\documentstyle[12pt,epsfig]{article}

\def\NPB{{\em Nucl. Phys.} B}
\def\PLB{{\em Phys. Lett.}  B}
\def\PRL{{\em Phys. Rev. Lett.}}
\def\PRD{{\em Phys. Rev.} D}
\def\MPL{{\em Mod. Phys. Lett.}  A}
\def\laq{\raise 0.4ex\hbox{$<$}\kern -0.8em\lower 0.62
ex\hbox{$\sim$}}
\def\gaq{\raise 0.4ex\hbox{$>$}\kern -0.7em\lower 0.62
ex\hbox{$\sim$}}

\def\L{{\cal L}}
\def\sframe{\hbox{$\underline{S\ }\!\!|$\ }}
\def\eframe{\hbox{$\underline{E\ }\!\!|$\ }}
\def\e{\hbox{\large $e$}}
\def\half{\hbox{\small $\frac{1}{2}$}}
\begin {document}
\titlepage
\begin{flushright}
BGU-PH-97/06 \\
\end{flushright}
\vspace{22mm}

\centerline
{\bf\Large Graceful Exit and Energy Conditions in String Cosmology}

\begin{center}
\vspace{10mm}
\centerline{Ram Brustein and Richard Madden}
\bigskip
\centerline{{\it Department of Physics, Ben-Gurion
University, Beer-Sheva 84105, Israel}}
\vskip 2  cm

{\large  Abstract} 

\end{center} 

\noindent
String cosmology solutions are examined in  a generalized 
phase-space including  sources representing arbitrary
corrections to lowest order string-dilaton-gravity effective action. 
We find a set of necessary conditions for a graceful exit transition from a
dilaton-driven inflationary phase to a radiation dominated era.
We show that sources allowing such a transition have to violate 
energy conditions similar to those appearing in singularity 
theorems of general relativity. 
Since familiar classical sources, excepting spatial curvature,
obey these energy conditions we conclude that a generic
graceful exit in string cosmology requires a new effective phase of matter. 
Our results clarify and generalize previous analyses and enable us
 to critically reexamine proposed non-singular cosmologies.

\vspace{5mm}

\vfill
\newpage

\setcounter{equation}{0}
\section {Introduction}

Duality symmetries of string cosmology equations
\cite{gv1,dual,tv},  suggest a mechanism \cite{gv1,gv2}  
for inflationary evolution. 
This mechanism is based on the fact that cosmological solutions to string 
dilaton-gravity come in duality-related  pairs, the  plus branch $(+)$, 
and the minus branch  $(-)$ \cite{bv}. The $(+)$ branch, even in the absence 
of potential energy, has inflationary solutions in which the Hubble parameter 
increases with time. The minus branch  $(-)$  can 
be connected smoothly to a standard Friedman$-$Robertson$-$Walker (FRW) 
decelerated  expansion of the universe with constant dilaton. 
The idea (so called ``pre-big-bang" scenario) is that evolution of
the universe starts from a state of very small curvature and coupling and
then undergoes a long phase of dilaton-driven kinetic inflation described by the
$(+)$ branch and at some later time joins smoothly standard  radiation dominated 
cosmological evolution, thus giving rise to a singularity free inflationary 
cosmology.

A key issue in the realization of this idea has been the nature of the
graceful exit transition from the  initial phase  of dilaton-driven
kinetic inflation to the subsequent standard radiation dominated evolution, in
particular, revealing  conditions under which such a  transition is allowed or
proving that under certain conditions a graceful  exit transition is forbidden.
In \cite{bv} it was argued,  and later proved \cite{kmo}, that such a transition
cannot occur while  curvature was below the string scale and the string coupling
was still weak, leading to the conclusion that an intermediate ``string phase"
of high curvature (previously  suggested as a possibility \cite {gv1,gv2})  or
strong coupling is actually required \cite{bggv}.  Recently, this was a subject
of intense activity, some investigations reinforcing and expanding the domain 
of validity of  the original conclusions \cite{reinforce},  some  resorting to
non-perturbative quantum effects to assist a transition \cite{quantumexit}, 
with varying degrees of success, while some have claimed that additional 
fields with certain form of interactions are required \cite{art,em,gmv}. 

Our approach is to use an effective description in terms of sources
that represent arbitrary corrections to the lowest order equations, depriving
us  the ability to obtain concrete solutions, but  allowing analysis of classes
of solutions and properties of effective sources. A solution  of the
``true equations"  necessarily satisfies the  effective equations,
even if the corrections to the lowest order equations are large, as long as the
whole framework does not break down.
There are different conformal frames  in which one can describe the 
equations. These are related by local field redefinitions which,  supposedly, 
do not affect physical observables \cite{gv3}. We  perform the analysis in the
string frame (\sframe) and in the ``lowest order Einstein frame" (\eframe).
We are able to relate necessary conditions for graceful exit to energy 
conditions appearing in singularity theorems of Eintein's general relativity
\cite{he},  and show that a successful exit requires violations of one of the
weakest of the energy conditions, the null energy condition (NEC).

Classical sources, such as 
perfect fluids with ``reasonable'' equation
of state, minimally coupled fields, etc, tend to obey NEC and therefore we 
conclude that a new kind of effective source is required to allow generic
graceful exit. We discuss various sources and comment about 
their relevance and we also put in context recent proposals offering 
non-singular evolution. We confirm our analysis by identifying the
NEC violating sources as resulting from special interaction terms between more
or less conventional sources.  None of the proposals seem to fit the full set of
requirements for a good string cosmology as defined in the text.

\setcounter{equation}{0}
\section {Effective string cosmology}

String theory effective action in 4 dimensions  takes the following form
\begin{equation}
S_{eff}^{{\sframe}}=
\int d^4 x \left\{ \sqrt{-g}\left[ \frac{e^{-\phi}}{16 \pi \alpha'}
\left(R+\partial_\mu\phi \partial^\mu\phi\right)\right]+
\half \L_m(\phi,g_{\mu\nu},...)\right\},
\label{effacts}
\end{equation}
where $g_{\mu\nu}$ is the 4-d metric and $\phi$ is the dilaton, 
the effective action is written here in the string frame (\sframe). 
The ``matter" Lagrangian $\L_m$ 
 may contain corrections to the lowest order 4-d action
 coming from a variety of sources, such as $\alpha'$ higher derivative
corrections, string quantum corrections, additional fields, denoted by ...,
 extra dimensions, string matter, etc.

We are interested in solutions to  the equations of motion derived from 
the action (\ref{effacts}) of the FRW type with vanishing spatial curvature
(non-vanishing spatial curvature may be included as an effective source in
the equations)
$ds^2= -dt_S^2+a_S^2(t) dx_i dx^i$ and $\phi=\phi(t)$. 
To allow solutions of this type the ``matter" energy-momentum 
tensor $T_{\mu\nu}=\frac{1}{\sqrt{-g}} \frac{\delta\L_m}{\delta g^{\mu\nu}}$, 
should have the form $T^\mu_{\ \nu}=diag(\rho,-p,-p,-p)$.

Of the three  equations of motion and one conservation equation
only three are independent and we choose the following as the independent set
\begin{eqnarray}
3H_S^2+\half\dot\phi^2-3 H_S\dot\phi&=& 
\half e^{\phi} \rho_S \label{n00eq} \\
\dot H_S +3 H_S^2 - H_S\dot\phi&=& 
\half e^{\phi}( p_S +\Delta_\phi\L_m)\label{hdoteq}\\ 
\dot\rho_S+3 H_S(\rho_S+p_S)&=& 
-\Delta_\phi\L_m \dot\phi, \label{nconseq}
\end{eqnarray}
where $\Delta_\phi\L_m=\half \frac{1}{\sqrt{-g}}  
\frac{\delta\L_m}{\delta\phi}$, $H_S=\dot a_S /a_S$, and we have fixed  our
units such that $16\pi\alpha'=1$. The constraint equation (\ref{n00eq}) may be
used to  rewrite equations 
 (\ref{n00eq}-\ref{nconseq}) in the form 
\begin{eqnarray}
\dot \phi &=& 3H_S\pm\sqrt{3H_S^2+e^\phi \rho_S} \nonumber\\
\dot H_S&=&\pm H_S\sqrt{3H_S^2+e^\phi \rho_S}+
\half e^\phi( p_S+\Delta_\phi\L_m) \nonumber\\
\dot\rho_S+3 H_S(\rho_S+p_S)&=& -\Delta_\phi\L_m \dot\phi.
\label{fstordm}
\end{eqnarray}
Our designation of the branches $(+)$ and $(-)$ corresponds to the sign
choice in these equations. Note that the constraint requires that 
$\e=3H_S^2+e^\phi \rho_S \ge 0$ ($\e<0$ is a classically forbidden
region) and that a change of branches can only
occur where $\e=0$. It is this surface in phase space that we refer to
as the ``egg'' \cite{bv,kmo}. Also notice that in the absence of strong sources
the $\dot H$ equation shows that the $(+)$ branch solutions are unstable
(${\rm sign}(H)={\rm sign}(\dot H)$), evolving to future singularities 
and conversely the $(-)$
branch will evolve from past singularities.

We may  perform a field redefinition to the ``lowest order
Einstein frame" (\eframe)  
$g_{\mu\nu}\rightarrow G_{\mu\nu}=e^{-\phi}g_{\mu\nu}$ and 
$\phi\rightarrow\varphi=\phi$ which transforms the lowest order 
dilaton-gravity action into Einstein's gravity and a minimally 
coupled scalar field, canonically normalized,
$
S_{eff}^{\eframe}=\int d^4 x \left\{ \sqrt{-G} \frac{R}{16 \pi G_N}
+\half \widetilde\L_m(\varphi,G_{\mu\nu},...)\right\},
$
where
$\widetilde\L_m(\varphi,G_{\mu\nu})=\L_m(\varphi,G_{\mu\nu}e^\varphi)
-\sqrt{-G} \partial_\mu\varphi \partial^\mu\varphi,$
so that we have included the dilaton kinetic energy in the definition
of the matter Lagrangian in  \eframe but not in \sframe.
Looking for solutions of the FRW type 
$ds^2= -dt_E^2+a_E^2(t) dx_i dx^i$ and $\varphi=\varphi(t)$,
the equations of motion reduce to the standard set, 
$H_E^2=\frac{1}{6} \rho_E$,..., etc.
(we have chosen units in which $16 \pi G_N=1$).
A particularly interesting linear combination of the {\small\it tt} 
and {\small\it ii} equations
which we will use later is the following
\begin{equation}
\dot H_E= -\frac{1}{4} (\rho_E+p_E).
\label{hdoteeq}
\end{equation}
We will need the relations between \sframe and \eframe quantities for later use
\begin{eqnarray}
dt_E&=&e^{-\phi/2} dt_S;\ a_E=e^{-\phi/2} a_S;\  
H_E=e^{\phi/2} (H_S-\half\dot\phi) \nonumber\\
p_E&=&e^{2\phi} p_S+\half e^{\phi}\dot\phi^2;\
\rho_E=e^{2\phi} \rho_S+\half e^{\phi}\dot\phi^2.
\label{estrans}
\end{eqnarray}

To construct a physically interesting cosmological evolution we require 
certain conditions on the asymptotic behaviour of solutions.
In our  setup we have to specify boundary conditions 
for four of the five functions \{$H_S,\phi,\dot\phi,\rho_S,p_S$\}\ 
and determine the boundary values of the remaining function subject to
the constraint (\ref{n00eq}) and branch choice.
The boundary conditions should guarantee that the problems 
of the standard model of cosmology are solved as in all models of inflation and 
that later evolution does agree with the observed universe 
(see also \cite{gv4}).
The initial conditions for $t\rightarrow -\infty$  
$H_S>0, \phi\ll 0,\dot\phi>0, \rho_S\simeq p_S\simeq 0$ and the choice of an
initial $(+)$ branch guarantee that  the solution  will emerge as an unstable
perturbation of the vacuum,  which can then evolve into a long period of dilaton
dominated kinetic inflation without reaching the very strong coupling regime
$\phi\gg 0$. 
The final phase we require to be ordinary radiation dominated FRW cosmology 
with a constant dilaton of intermediate value, 
$H>0, \phi\simeq 0,\dot\phi=0, \rho= 3p$. 
Note that in this case \sframe and \eframe quantities are proportional to each
other  and we therefore drop the subscript $S$. The final
state of radiation, for which $T^{\mu}_{\ \mu}=\rho-3p=0$, 
is consistent with a constant decoupled dilaton,
which in turn implies that we are on a $(-)$ branch. 
Thus at least one branch change event 
from $(+)$ to $(-)$ has to occur in the course of the evolution.

In addition, we require that throughout 
the evolution,  even during the string phase, all components of curvature 
invariants,  $\phi$, $\rho$, $p$, and their derivatives remain finite. 
We  do not allow a singular field redefinition
for which new curvature invariants derivatives and sources are finite. The 
reason for this requirement is to guarantee that there are some physical 
observables which survive the evolution. This requirement is similar in spirit 
to the bounded curvature idea in \cite{bm}, and less similar 
to ideas about smoothing and resolution of singularities suggested 
in string theory context \cite{topchange}. 

Finally we note that our definition of the ``lowest order Einstein frame''
(\eframe)  coincides with the true Einstein frame with decoupled dilaton and
gravity near our initial and final conditions.
In some of the  models we examine, we find
that \eframe does not correspond to what might be considered a true Einstein
frame, since dilaton-gravity couplings have been absorbed into the effective
$\rho$ and $p$.

\setcounter{equation}{0}
\section {Branch change and exit}

At least one branch change event 
from $(+)$ to $(-)$ has to occur in the course of the evolution, as explained 
in the previous section. We focus our attention on the last branch change which
has to be from $(+)$ to $(-)$. Recall
 the egg variable ${\e}= 3H_S^2+ \rho_S e^\phi$. 
A necessary condition for a branch change to occur is that the 
evolution of the solution touch the egg, the surface in phase space
defined by $\e=0$, which requires that 
$\rho_S<0$ in some region of parameter space. Continuity of the
solution then forces  a branch change.  The region of negative $\rho_S$ is
divided into two parts, $H_S>0$ (above the egg) and $H_S<0$
(below the egg) separated by the classically forbidden region $\e<0$. 
Notice that above the egg $\dot \phi>0$ and below the
egg $\dot \phi<0$.

Using the equations of motion we can show that
\begin{equation}
\dot{\sqrt{\e}}= \pm \left[3H_S^2+ \half \rho_S e^\phi-
\half  \Delta_\phi\L_m e^\phi\right].
\label{eggeqphi}
\end{equation}
For an immediate application of eq.(\ref{eggeqphi}) consider a  
$(+)$ solution approaching 
the egg. For it to hit the egg $\e$ must decrease to zero, 
so $\dot{\sqrt{\e}}<0$ just before the egg is hit.
If the sources are not coupled to the dilaton, then $\Delta_\phi\L_m=0$
and we can write 
$\dot{\sqrt{\e}}=3H_S^2+ \half \rho_S e^\phi=\e-\half e^\phi \rho_S$.
But this quantity is positive just before an egg hit, since we have 
observed $\rho_S<0$ there. In this case the egg repels $(+)$ branch solutions
and we conclude branch change is not possible.
The matter sources have to be rather strongly 
coupled to the dilaton for a branch change to be possible.

We proceed to derive a further ``global" necessary condition for exit 
based on the requirement that the last branch change will really be 
the last branch change. Recall that if the evolution of the solution touches the
egg, a branch change must occur. We therefore require that the evolution 
of the solution does not touch the egg anymore after the last branch change.
It is enough to make sure that no further branch changes occur 
until the ``moment of escape", after which $\rho_S$ is always positive and
an additional branch change is impossible. We discuss the generic situation, 
deferring discussion of possible pathological cases as in \cite{kmo} to
another opportunity.
To derive the global condition, following \cite{kmo}, we  use eq.(\ref{hdoteq}) 
to eliminate $\Delta_\phi\L_m$ from eq.(\ref{eggeqphi})
\begin{equation}
\dot{\sqrt{\e}}= \pm \left[-\dot H_S+H_S\dot\phi+ 
\half  e^\phi (\rho_S+p_S)\right].
\label{eggeq}
\end{equation}
Assuming a successful escape, we integrate eq.(\ref{eggeq}) from the moment 
of branch change $t_{\rm hit}$, at
which $ {\e}=0$ and $\dot\phi=3 H_S$ to the moment of escape $t_{\rm escape}$, 
at which $\rho_S=0$, $\e=\sqrt{3} H_S$ and $\dot\phi=(3-\sqrt{3}) H_S$
\begin{equation}
\int\limits_{t_{\rm hit}}^{t_{\rm escape}} dt\left[ -\dot{\sqrt{\e}}
+\dot H_S\right]=
\int\limits^{\phi(t_{\rm escape})}_{\phi(t_{\rm hit})} H_S d\phi+
\int\limits_{t_{\rm hit}}^{t_{\rm escape}} dt\ \half  e^\phi (\rho_S+p_S).
\label{escape1}
\end{equation}
Performing the integration on the left-hand-side we obtain
\begin{equation}
-\left[(\sqrt{3}-1)H_S(t_{\rm escape})+H_S(t_{\rm hit}) +{\cal A} \right] =
\int\limits_{t_{\rm hit}}^{t_{\rm escape}} dt\ \half  e^\phi (\rho_S+p_S),
\label{escape2}
\end{equation}
where ${\cal A}=\int\limits^{\phi(t_{\rm escape})}_{\phi(t_{\rm hit})} H_S d\phi$ is
the positive  area of the domain bounded by the solution
 and $\phi$-axis shown in figure 1.
Note that the left-hand-side of eq.(\ref{escape2}) is negative. In fact, 
it probably has a rather 
large magnitude because $H_S$ is likely to be rather large throughout this epoch.
We conclude that a necessary condition for a successful escape is that 
\begin{equation}
\int\limits_{t_{\rm hit}}^{t_{\rm escape}} dt\  e^\phi (\rho_S+p_S)<0.
\label{sec}
\end{equation}
For some specific cases this condition was derived in \cite{kmo}.

A global condition in \eframe is obtained as follows. 
We may show, using relations (\ref{estrans})
and the condition $\e=0$, that the value of $H_E$ at 
the moment of ``egg-hit" is negative 
\begin{eqnarray}
H_E(t_{\rm hit})&=&e^{\phi(t_{\rm hit})/2} \left(H_S(t_{\rm hit})-
\half\dot\phi(t_{\rm hit})\right) \nonumber\\
&=&e^{\phi(t_{\rm hit})/2}\left (H_S(t_{\rm hit})-
\frac{3}{2}H_S(t_{\rm hit})\right) \nonumber\\
&=&-\half e^{\phi(t_{\rm hit})/2} H_S(t_{\rm hit})<0.
\end{eqnarray}
We may also show, using 
 relations (\ref{estrans}) and the condition $\e=\sqrt{3} H_S$, that the value of 
$H_E$ at the moment of escape  is positive 
\begin{eqnarray}
H_E(t_{\rm escape})&=&e^{\phi(t_{\rm escape})/2} \left(H_S(t_{\rm escape})-
\half\dot\phi(t_{\rm escape})\right)\nonumber \\
&=&e^{\phi(t_{\rm escape})/2}\left(H_S(t_{\rm escape})-
\frac{3-\sqrt{3}}{2}H_S(t_{\rm escape})\right)\nonumber \\
&=&e^{\phi(t_{escape})/2}\frac{\sqrt{3}-1}{2}H_S(t_{escape})>0.
\end{eqnarray}
\begin{figure}
\begin{center}
\hspace{-1.0in}\epsfig{file=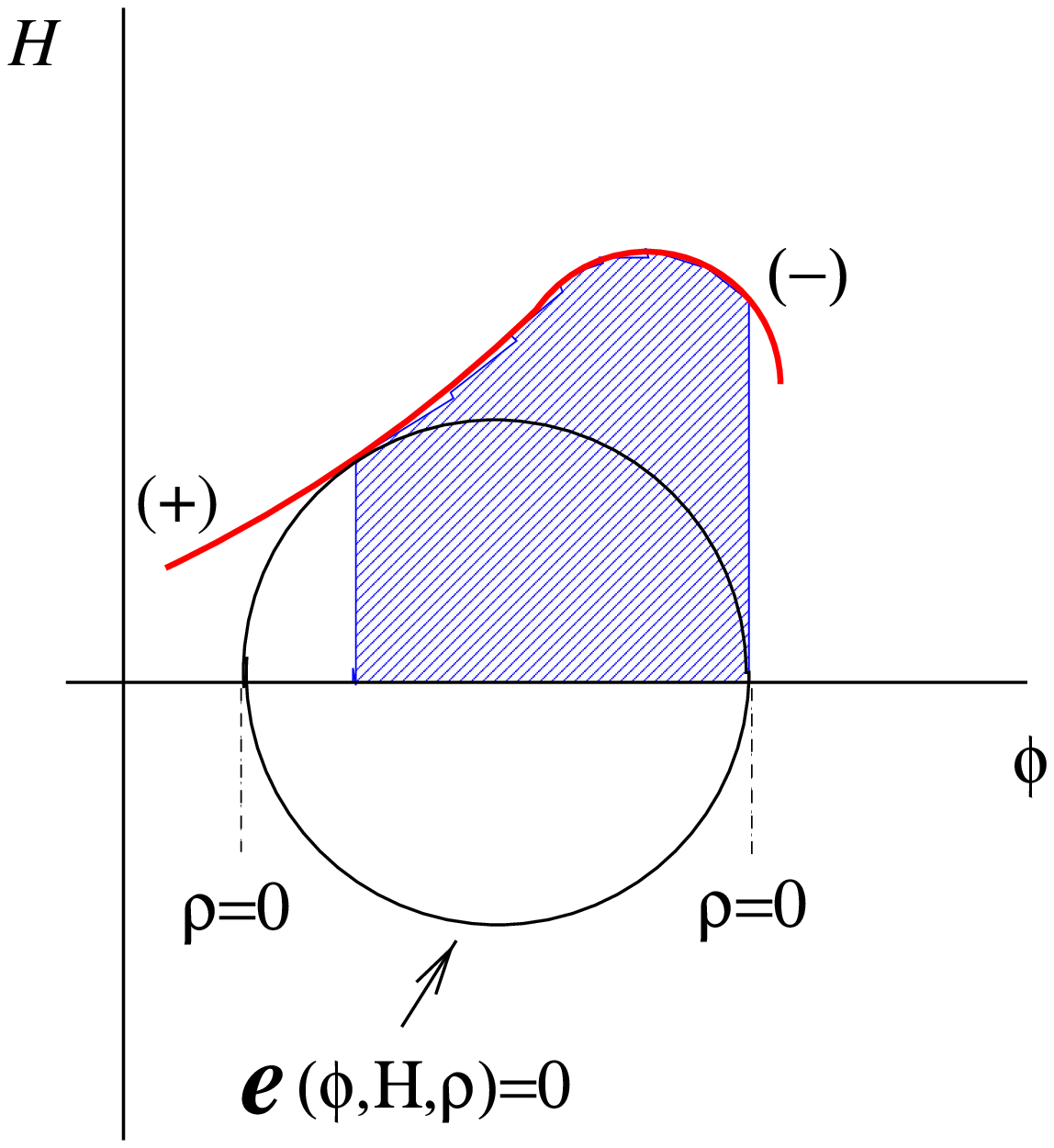, height=5cm, bbllx=1pt, bblly=1pt,
bburx=328pt, bbury=358pt}
\epsfig{file=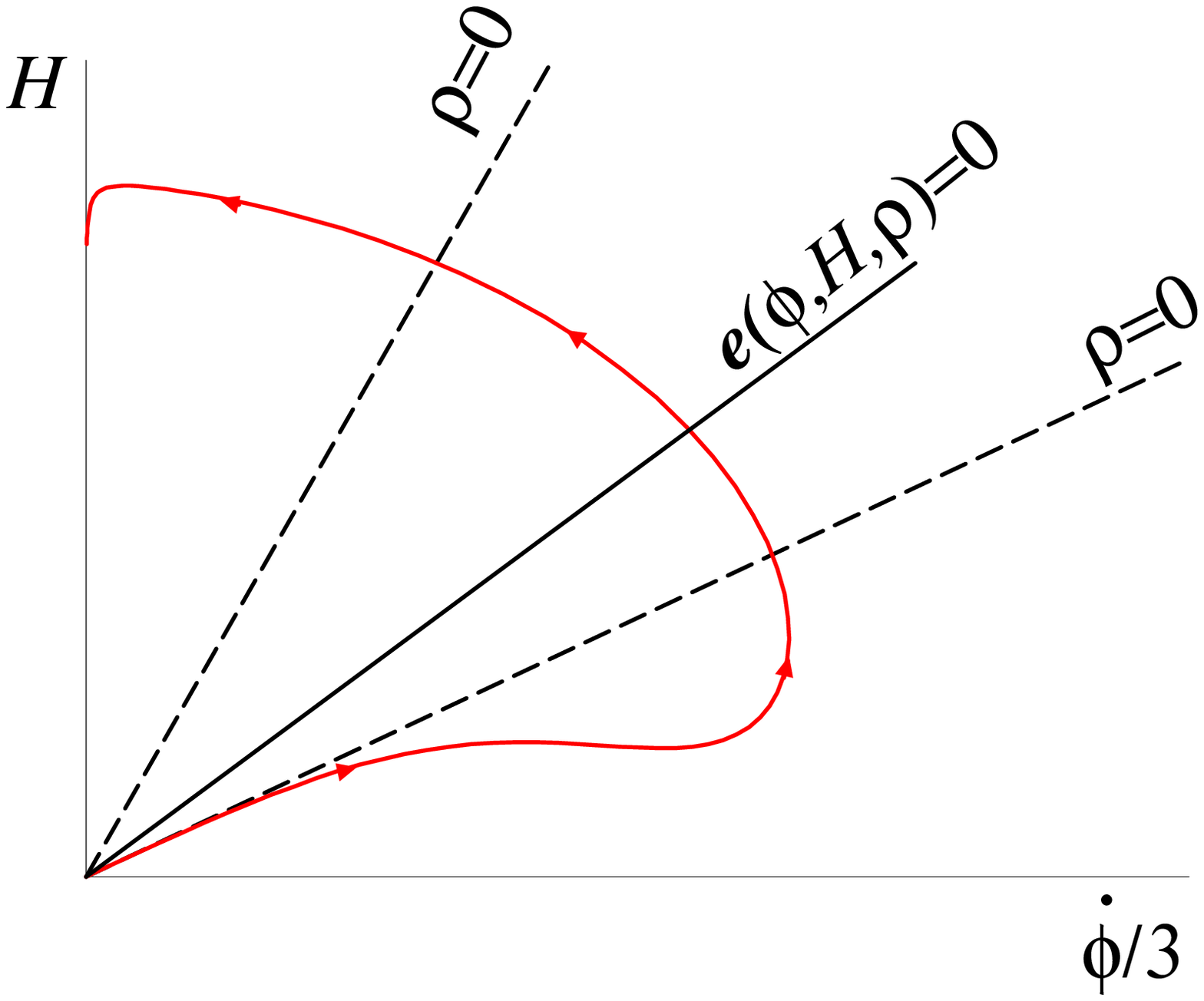, height=4.5cm, bbllx=1pt, bblly=1pt,
bburx=328pt, bbury=358pt}
\end{center}
\caption{
\small\it
\label{f:egg}
Shown are two (hypothetical) examples of a complete graceful exit in \sframe. 
The one on the left is shown in $(\phi,H)$ plane with an arbitrary origin
for the $\phi$ axis. 
A transition from $(+)$ branch to $(-)$ branch
occurs when the solution hits the curve $\e(\phi,H,\rho)=0$ (the egg) 
and is then followed by a successful escape to a point where $\rho_S=0$.
The shaded area 
${\cal A}=\int H d\phi$  
is always positive since $\phi$ is increasing and $H>0$.
The one on the right is shown in the $(\dot\phi,H)$ plane with the 
axes origins at $(0,0)$. The solution begins asymptotically near 
the lower dashed line which is $\rho=0$ (vacuum) for the
$(+)$ branch. The solution then crosses this line developing $\rho<0$
and makes a branch change crossing the solid line $\e=0$. Finally it
completes a graceful exit to the $\rho>0$ region 
by crossing the upper dashed line which
is $\rho=0$ for the $(-)$ branch and reaches  $\dot \phi=0$
 and radiation domination.} 
\end{figure}
Therefore the total change in $H_E$ is positive
$
\Delta H_E=\int\limits_{t_{\rm hit}}^{t_{\rm escape}} dt_E\  \dot H_E>0,
$ which implies, using eq.(\ref{hdoteeq}), that 
$
-\frac{1}{4}\int\limits_{t_{\rm hit}}^{t_{\rm escape}} dt_E\  (\rho_E+p_E)>0.
$
We conclude that a necessary condition for a successful escape is that 
\begin{equation}
\int\limits_{t_{\rm hit}}^{t_{\rm escape}} dt_E\  (\rho_E+p_E)<0.
\label{eec}
\end{equation}
Soon we will relate energy conditions (\ref{sec}), (\ref{eec}) to those used 
in singularity theorems of general relativity and discuss their significance, 
but for now we  would like to discuss the relationship of a bounce in 
\eframe to  branches in \sframe.

A bounce in \eframe is a transition from a contracting to an expanding universe
$H_E<0\Rightarrow H_E>0$. We wish to refine the relationship between a 
bounce in \eframe and a branch change and exit. Recall that
$H_E=e^{\phi/2}(H_S-\half\dot\phi)$ and that for  $(+)$ branch $\dot\phi-3H_S>0$ 
and for  $(-)$ branch $\dot\phi-3H_S<0$.
Considering  solutions for which $H_S>0$ always, 
we obtain the following classification
\begin{eqnarray}
H_E\!&<&\!0 \Rightarrow 
\cases{
 \dot\phi>3H_S & $\Rightarrow$ (+)\ branch  \cr
2H_S<\dot\phi<3H_S & $\Rightarrow (-)$\ branch. \cr 
}\nonumber \\
H_E\!&>&\!0\Rightarrow (-)\ \hbox{\rm branch}. 
\end{eqnarray}
 Conversely, (always keeping
$H_S>0$) 
\begin{eqnarray}
(+)\ \hbox{\rm branch}&\Rightarrow&  H_E<0 \nonumber \\
 (-)\ \hbox{\rm branch} &\Rightarrow&
\cases{
 \dot\phi<2 H_S & $\Rightarrow H_E>0$  \cr
2H_S<\dot\phi<3H_S & $\Rightarrow H_E<0.$  \cr 
}
\end{eqnarray}
The conclusion is that a bounce does not necessarily imply a branch change
because a bounce can occur while the solution remains
a $(-)$ branch solution, and that
a branch change does not necessarily imply a bounce because we have
seen that $H_E$ is negative in the region around the branch change
point. The bounce is only guaranteed if the solution proceeds to exit.\\

Let us summarize now the set of necessary conditions for 
implementing successful exit
\begin{itemize}
\item
Initial conditions of a (+) branch and $H_S,\dot\phi>0$  require
$H_E<0$.
\item
A branch change from (+) to $(-)$ has to occur while $H_E<0$. At that point
$\e=3H_S^2+\e^{\phi}\rho_S$ has to vanish and therefore $\dot{\e}$ just before
the branch change has to be negative, and positive just after.
\item
A successful escape and exit completion requires a bounce in \eframe 
after the branch change has occurred, ending up with $H_E>0$.
\item
Further evolution is required to bring about a radiation
dominated era in which the dilaton effectively 
decouples from the ``matter" sources.
\end{itemize}

We repeat for emphasis that a bounce in \eframe is not  a sufficient condition 
for an exit, and as already
shown in \cite{bv,kmo} a branch change is not a sufficient condition
for an exit. We also state the trivial fact that  necessary conditions 
are not sufficient  conditions and therefore even if they are satisfied
they do not ensure successful exit transition.

\setcounter{equation}{0}
\section {Energy conditions}

We wish to relate the energy conditions in eqs.(\ref{sec}, \ref{eec})
to those that appear in singularity theorems of general relativity \cite{he}. 
The energy conditions that are most relevant to the
present discussion are the local null energy condition (NEC) and its global 
version, the averaged null energy condition (ANEC) \cite{anec1}, a subject of 
some recent  investigations \cite{anec}. The NEC requires that  
$T_{\mu\nu} k^\mu k^\nu \ge 0$ for all null vectors $k^\mu$ and
the ANEC requires that 
$
\int\limits_\gamma T_{\mu\nu} k^\mu k^\nu d\lambda \ge 0,
$
where $\gamma$ is a null curve, $k^\mu$ is a tangent (null) vector and $\lambda$ is a 
generalized affine parameter.
For $T^\mu_{\ \nu}=diag(\rho,-p,-p,-p)$ the NEC reduces to 
$\rho+p \ge 0$ and the ANEC reduces to 
$\int\limits_\gamma (\rho+p) d\lambda \ge 0.$ The NEC is one of the weakest
energy conditions and its violation implies 
a violation of the weak, dominant, strong, and generic energy conditions.

In \eframe, condition (\ref{eec}) states 
that a necessary condition for an exit transition is that the
NEC is violated for some range of time. In \sframe condition
(\ref{sec}) similarly implies that $\rho_S+p_S< 0$ for some range of 
time.
In  \sframe $\rho+p$ can be negative even if in \eframe both $\rho$ and $p$
are positive $\rho_E+p_E=e^{2\phi}(\rho_S+p_S)+ e^\phi \dot\phi_S^2$,
therefore condition (\ref{eec}) seems to be a more restrictive condition.  

Our conclusion is that the NEC has to be violated to allow a 
graceful exit transition, implying that an effective source which
enables graceful exit must invalidate the conditions under which
 the classical singularity theorems of general relativity were proved,
as expected since we now have a non-singular cosmology.
Energy conditions are arguably the weakest link in the chain of 
arguments leading to singularity theorems, in particular, we are not 
aware of any general argument that forbids from first principles NEC 
violating sources. 
In fact, violations of NEC are  expected to occur due to quantum effects, as
in the celebrated example of Hawking radiation of black holes, 
although their strength in some specific cases is under debate \cite{anec}, and 
spatial  curvature is a classical effective source that violates NEC (see below).
Energy conditions in \sframe were also discussed in a different context
\cite{rbp}.

\setcounter{equation}{0}
\section {Sources}

We now turn to discuss possible sources and their role in graceful exit 
transition, paying some attention to suggestions in the literature.

First, let us point out that our results generalize
and reinforce the conclusions of \cite{bv,kmo,reinforce} that found
that adding fields and sources ``off the shelf" (including all 
types of stringy matter) will not facilitate graceful
exit since they obey NEC.  Also, attempts to tinker with minimally coupled fields 
to produce a negative $\rho+p$ tend to create ghosts 
or tachyons, unacceptable if classical
stability and quantum unitarity are required. 
Obviously, other kinds of effective sources are necessary.

Effective sources that naturally appear in string theory are 
\begin{itemize}
\item
 Higher derivatives classical ($\alpha'$) corrections. Because of technical
problems related to on-shell conditions, string theory is not very specific 
about the coefficients of  higher derivative terms, and in particular, 
at the moment,
there is no stringy principle able to constraint the effective $\rho+p$ of these
terms. As we discuss below, in some cases they do provide negative $\rho+p$.
\item
Quantum corrections which manifest themselves
as higher derivative corrections  with a specific dilaton dependence,
or in modification of the form of 
the  coupling of the dilaton and gravity to other fields.
It may well be possible to devise  ad-hoc modifications that will allow
an exit transition. However, it turns out that large classes of models,
including most of those that appeared in the literature have some
property that prevents graceful exit from happening.
In a different context, It has been argued \cite{anec} 
that quantum NEC violations are generic, however, their strength in various 
cases is under debate. Note that Hawking radiation of black holes
allowing the area of event horizons to shrink is a prime example of such 
a quantum effect.

A more radical departure from our framework consists of allowing Euclidean
solutions describing classically forbidden tunneling transitions. 
We have nothing
to say about this possibility, clearly outside the scope of our 
current investigations.

\item
 Extra dimensions. 
As already discussed in \cite{bv} the effective contribution of simple
contracting or expanding extra dimensions is positive to both $\rho$ and
$p$, making them a graceful exit suppressing source.
\item
Spatial curvature is perhaps the most obvious source with negative $\rho+p$.
Recall that in \eframe $\rho_{\rm curv}=-6k/a^2(t)$ and 
$p_{\rm curv}=2k/a^2(t)$ where $k=0,\pm 1$, and therefore 
$\rho_{\rm curv}+p_{\rm curv}=-4k/a^2(t)$. Thus, for $k=+1$, 
$\rho_{\rm curv}+p_{\rm curv}<0$. Combining spatial curvature with
another off the shelf source, a cosmological constant yields cosmologies
of the general form $a(t)=\cosh(t)$ with the negative $\rho+p$ from
the curvature doing the job of reversing the contracting deSitter
solution to the expanding one.
We do have some reservations about using
spatial curvature in this context since one of the jobs 
that a dilaton-driven inflationary phase
does best is to erase initial spatial curvature. So if the universe
successfully inflates we don't expect curvature to be a strong source
when its help is required for graceful exit.
\end{itemize}

In \cite{bggv}, following the conclusions of \cite{bv,kmo} the necessity of
an intermediate string phase was recognized. The simplest possible 
extrapolation
of the solution was suggested, $H_S=const, \dot\phi=const$. The simple
extrapolation was taken up more seriously  in \cite{gmv}, showing that indeed
there is an all order solution of  this type and furthermore that this ``simple
string phase" can be reached continuously from the $(+)$ branch initial
conditions. Our results show that it is impossible to complete a graceful exit
transition during a simple string phase. Recall from relations (\ref{estrans})
that $H_E=e^{\phi/2} (H_S-\half\dot\phi)$, and therefore
if $H_S$ and $\dot\phi$ are constants, the sign of $H_E$ cannot change
during this phase.
An examination of the numerical results of \cite{gmv} (for d=3)
 reveals that a branch change induced by some higher derivative terms has
indeed occurred, but that $H_E$ is negative  throughout and that a simple string
phase solution represents a singular collapse in a finite time in \eframe. As
the authors point out, with the still growing dilaton, quantum
corrections may aid the transition to a completed graceful exit.

Other analyses \cite{art,em}, presenting non-singular cosmologies
do not use the desired boundary conditions which degrades somewhat their 
relevance.
In \cite{art} $H_E>0$ throughout the whole evolution, and therefore
 their evolution cannot exhibit a branch change, indeed, their solution is 
always a  $(-)$ branch solution. In \cite{em} the initial conditions are such
that $H_S<0, \dot\phi<0$ in the $(+)$ branch and ``final conditions" 
$H_S>0, \dot\phi<0$ in the $(-)$ branch. 
However, the solution in \cite{em} does  exhibit branch 
changes and bounces, the Gauss-Bonnet with its specific coupling
to the extra modulus field supplying the effective NEC violating source. 
Further examination also shows  the dilaton has a very little influence in
driving the system. Asymptotically these solutions are dominated by 
terms coming from the spatial curvature and the coupling between
the modulus and Gauss-Bonnet terms so initial conditions are nowhere near the
dilaton-gravity perturbative vacuum, making these solutions 
irrelevant to the envisioned graceful exit transition.

Finally, we look at examples of non-singular cosmologies without
specifically stringy motivations \cite{rama}. These
can be constructed from singular Einstein frame evolutions by choice of an
appropriate conformal scaling, which dictates the non-minimal coupling for the
scalar. Again, this coupling produces NEC violating sources in \sframe and 
\eframe but, of course, not in the true Einstein frame. So, while appearing 
non-singular, they cannot solve the problems addressed by inflation such as
the horizon problem \cite{twlf}.

\setcounter{equation}{0}
\section{Conclusion}

A singularity free cosmology requires the violations of the classic singularity
theorems of general relativity. We have found that a graceful exit transition 
from a dilaton-driven inflationary phase to a radiation dominated era
that allows a non-singular string cosmology must be induced by an effective
source that violates one of the weakest  energy conditions used in these
theorems, the null energy condition (NEC). We examine various proposed
non-singular cosmologies and find NEC violating sources induced by different
forms of interactions, but conclude that none of these are directly
relevant to  graceful exit in string cosmology. 

We are unaware of any fundamental physics principle that forbids or requires
the existence of  NEC violating sources.  Although most known classical
sources do obey NEC and therefore cannot induce an exit,  spatial curvature
is effectively an NEC violating source and there is evidence that 
quantum mechanical violations of NEC are generic and related to 
trace anomalies, a common quantum effect. A prime example of a quantum phenomenon 
that violates NEC is Hawking radiation of black holes that allows the
shrinking of the area of event horizons, even though this is forbidden by 
singularity theorems. With this in mind, we feel that the need for
violations of NEC to accomplish the graceful exit should not be taken as 
evidence that it is a forbidden transition, but rather that it may involve
different physics than expected. \\

\noindent
{\Large\bf Acknowledgment}\\

This research is supported in part by the Israel Science Foundation 
administered by the Israel Academy of Sciences and Humanities.
We thank Howard Burton and Nemanja Kaloper for discussions about \cite{rama}.

\end{document}